\documentclass{article}

\usepackage{arxiv}

\usepackage[utf8]{inputenc} 
\usepackage[T1]{fontenc}    
\usepackage{hyperref}       
\usepackage{url}            
\usepackage{booktabs}       
\usepackage{amsfonts}       
\usepackage{nicefrac}       
\usepackage{microtype}      
\usepackage{lipsum}
\usepackage{graphicx}
\graphicspath{ {./images/} }

\title{AI agents for facilitating social interactions and wellbeing}

\author{
 Hiro Taiyo Hamada \\
  Araya Inc.\\
  Akasaka, Tokyo, Japan \\
  \texttt{hamada\_h@araya.org} \\
   \And
 Ryota Kanai \\
  Araya Inc.\\
  Akasaka, Tokyo, Japan \\
  \texttt{kanair@araya.org} \\
}

\begin{document}
\maketitle
\begin{abstract}
Wellbeing AI has been becoming a new trend in individuals' mental health, organizational health, and flourishing our societies. Various applications of wellbeing AI have been introduced to our daily lives. While social relationships within groups are a critical factor for wellbeing, the development of wellbeing AI for social interactions remains relatively scarce. In this paper, we provide an overview of the mediative role of AI-augmented agents for social interactions. First, we discuss the two-dimensional framework for classifying wellbeing AI: individual/group and analysis/intervention. Furthermore, wellbeing AI touches on intervening social relationships between human-human interactions since positive social relationships are key to human wellbeing. This intervention may raise technical and ethical challenges. We discuss opportunities and challenges of the relational approach with wellbeing AI to promote wellbeing in our societies.
\end{abstract}


\section{Introduction}
\label{sec:intro}
COVID-19 has revealed the importance of the sense of belongingness and loneliness in mental health of our societies \cite{COVID-192021}. Wellbeing has attracted attention of psychology and public health for improving the mental health of individuals and organizations and has become one of the main targets for public health organizations such as the World Health Organization (WHO) \cite{Topp2015}.
Wellbeing has been studied intensively in the context of psychology \cite{Andrews1976, Diener2018, Topp2015}. In psychology, multiple constructs of wellbeing have been proposed \cite{Dodge2012}. For example, Ryff and Keyes proposed that wellbeing is composed of multiple factors such as autonomy, environmental mastery, personal growth, positive relations with others, purpose in life, and self-acceptance \cite{Ryff1995}. Although there are differences in emphasis among psychological theories, positive social relationships have been identified as a crucial factor.

There is a new trend to apply artificial intelligence (AI) to enhance wellbeing due to the development of emotion analysis technologies such as genome-wide analysis, computer vision, and natural language processing (NLP). Multiple services have been introduced to analyze or intervene in mental health by accessing peoples' emotions. For example, analyses by personal genetic data \cite{Fang2020}, images \cite{Reece2017}, and texts in social media \cite{Chancellor2020} predict risks and mental conditions including mental disorders. Some applications further intervene in mental conditions based on theories of psychological intervention such as cognitive-behavioral therapies (CBTs)\cite{vanAgteren2021}. Although social factors are known to be crucial, most AI applications for wellbeing focus on individuals and much less on social groups. Given that we spend most of our time in multiple social groups such as family, workplaces, schools, social clubs, etc., the opportunities and potential impact of such group-targeted AI applications would be enormous. However, AI applications of social groups for wellbeing have attracted little attention.

Here, we present an overview of the emergent role of AI-augmented agents for social interactions. First, we investigate the literature on psychological wellbeing and provide a two-dimensional classification of AI-augmented agents: individual/group and analysis/intervention. The first dimension concerns whether wellbeing AI is used for the analysis or the intervention. The second dimension focuses on whether an AI-augmented agent targets individuals or groups. We point out opportunities for the recently emerging approach, the so-called relational approach, where AI-augmented agents are applied to human-human interactions within groups. Finally, we discuss challenges in the relational approach of AI-augmented agents. We shed light on broader opportunities for AI-augmented agents, and highlight technological and ethical challenges for promoting wellbeing in the real and virtual societies.

\section{Social Construct of Wellbeing}
\label{sec:social_construct}
The notion of wellbeing has attracted attention in the context of healthy individual lives and societies. Subjective wellbeing (SWB) has been widely measured as a screening tool for mental disorders based on self-reported questionnaires such as the WHO questionnaire. Several models of SBW have been proposed \cite{Andrews1976, Dodge2012, Topp2015}. SWB is composed of multiple facets comprising two components \cite{Schimmack2008,Luhmann2012}: affective and cognitive evaluations of one's life. The affective evaluation measures the emotional experiences of people in daily lives while the cognitive evaluations measure how people evaluate their lives based on their ideals. The affective and cognitive aspects are associated with different scales such as daily emotional experience and life satisfaction, respectively \cite{Diener2018}. Recent studies also suggested that another supplementary factor, harmony in life in a social context is also associated with SWO \cite{Kjell2016}. Harmony in life reflects social and environmental situations and is associated with psychological balance and flexibility in life. Therefore, social factors play a critical role in SWO.

Social personalities for wellbeing have been widely studied, showing consistent results. A recent meta-analysis, for example, revealed that widely used personality factors (e.g. NEO-PI-R and HEXACO questionnaires) are correlated with several aspects of SWO such as life satisfaction, positive/negative affect, and positive relation with others \cite{Anglim2020}. The study especially found that these aspects of SWO are positively correlated with extraversion and conscientiousness although negatively correlated with neuroticism. The sensitivity of SWO, thus, could reflect the personality traits of subjects. It is noteworthy that extraversion, as well as neuroticism and conscientiousness, also influence related factors like depressive symptoms \cite{Hakulinen2015}. Extraversion is a social indicator for higher positive relationships with others. Meanwhile, neuroticism is another social indicator for less positive relationships related to loneliness \cite{Buecker2020}. The association between these personalities and wellbeing-related factors supports the idea to promote wellbeing via positive relationships.

It is an important question whether behavioral practice can change social relationships and wellbeing. Multiple attempts showed enhancement of wellbeing as well as associated factors by healthy behaviors such as exercising \cite{Chekroud2018} and psychological interventions \cite{vanAgteren2021}. A cross-sectional study from 1.2 million individuals in the U.S. showed that physical exercising routines such as popular team sports, aerobic, and gym activities decrease up to 22\% of mental health burdens compared to the non-exercising group \cite{Chekroud2018}.

Furthermore, different psychological interventions such as behavioral activation interventions (BA), positive psychological intervention (PPI), and mindfulness-based interventions (MBI) also showed small-to-moderate effects on wellbeing \cite{vanAgteren2021}. Social interventions alleviated social isolation \cite{Dickens2011} and loneliness \cite{Masi2011}. These empirical findings further support that interventions including exercises and psychological intervention for social relationships on wellbeing can promote wellbeing and prevent mental illness.

We spend many hours with family, friends, and colleagues. Subjective wellbeing in groups such as working place is also studied well \cite{Harter2003, Jain2009}. Working environments and social networks influence wellbeing and healthy behaviors. Associations between work environment and wellbeing are known \cite{Harter2003,Bowling2010}. Life satisfaction and other related factors such as job satisfaction and positive affect are related to wellbeing. Another evidence further showed that mindfulness training had small-to-moderate effects on psychological distress, wellbeing, and sleep \cite{Bartlett2019} although the influence on work performance could not be concluded due to the insufficiency of pooled data. Internet-based interventions on workers showed small-to-moderate effects on work effectiveness and psychological wellbeing in workplaces \cite{Carolan2017}.

Psychological interventions on social networks, so-called social network interventions, are also effective on wellbeing \cite{Hunter2019}. This relatively new approach cares for changes in information flow by intensifying, deleting, and transferring social ties \cite{Valente2012}. The social network intervention is expected to enhance the effectiveness of health outcomes such as lower drug use, healthy sex behaviors, stronger social support, and wellbeing.

It is interesting to ask whether this approach is useful for social media and online gameplay. There is a strong public interest in the association of social media use and game playing with mental health. Their potentially harmful influences on mental health have often drawn public attention \cite{Huang2010, Prescott2018}, but the relationship remains unclear, perhaps due to huge differences in design and concepts within social media \cite{Sakurai2021} and games \cite{Johannes2021}. Communication within online video games such as e-sports can be essential for effective team performance. Effective social intervention may increase not only team performance and wellbeing, but the potential of such social interventions remains clear. Findings on group wellbeing, nonetheless, reveal another potential target of interventions in our societies.

To sum up, existing literature revealed associations of social factors related to genetics, environments, and behaviors with wellbeing. These multiple findings clarify possibilities of interventions of subjective wellbeing as well as group wellbeing.

\section{Types of Wellbeing AI}
The effectiveness of interventions on wellbeing triggered expectations to conduct research and development along with a trend of digital therapeutics. Digital therapeutics are evidence-based therapeutic interventions with software programs to cure and prevent medical disorders. There is also a trend to apply AIs to such evidence-based interventions in mental health \cite{DAlfonso2020}. In this section, we summarize and explain the types of AI-augmented applications such as robots, avatars, and bots for wellbeing. By doing so, we clarify currently active approaches in wellbeing AI.

We categorize AI applications for wellbeing with two-dimensional axes: analysis/intervention and individual/group (Table.\ref{tab:table1}). The first dimension means whether the aim of digital wellbeing is analysis or intervention. Many potential applications focus on monitoring emotional states or related factors associated with wellbeing through genes \cite{Fang2020}, images \cite{Canedo2019}, and texts \cite{Chancellor2020} from social media.

Other applications target the individual status of wellbeing-related factors through apps. Meanwhile, positive social networks are a crucial social basis for groups as well as individuals. There are only a few applications of wellbeing targeting groups from this perspective \cite{Narain2020}. For example, a Facebook messenger chatbot, Sunny, is meant to promote social interactions and wellbeing within groups \cite{Narain2020}. We belong to multiple social groups in different contexts such as schools, workplaces, sports clubs, and our houses. Scopes of group wellbeing should be also broad. There are some studies on group wellbeing and proposals to promote such group wellbeing from AI-augmented agents such as robots \cite{Kim2020,Shin2021,Tennent2019, Traeger2020}. However, group wellbeing targeted by AI applications is relatively understudied and may bring new opportunities for the promotion of wellbeing.

In summary, most of the applications in wellbeing AI focus on analysis and psychological interventions of individual wellbeing through mobile devices. Meanwhile, group wellbeing is relatively under-explored but may have a huge impact on our societies.

\section{The Relational Approach for Group Wellbeing with AI}
In this section, we overview a relational approach for group wellbeing with the literature of analysis and interventions on human-human interactions with AI agents. By doing so, we outline opportunities of wellbeing AI for group wellbeing. We then discuss types of the relational approach: analysis of group dynamics and social connectedness. Finally, the challenges of the relational approach will be discussed.

\subsection{Literature Review on Analysis and Interventions for Social Groups with AI}

\begin{table}
  \centering
  \begin{tabular}{lllll}
    \toprule
    \midrule
    Individual/Group & Analysis/Intervention & Categories & Examples \\
    \midrule
    Individual & Analysis & Genetics & - Depression Risk \cite{Fang2020}     \\
     &  & & - Subjective Wellbeing \cite{Roysamb2019}     \\
     &  & & - Social Support \cite{Wang2017}    \\
\cmidrule(r){1-4}
    Individual & Analysis & Mental Health & - Emotion Detection \cite{Canedo2019}\\
     &  & & - Screening Mental Health Status \\
     &  & & on Social Media \cite{Chancellor2020}\\
\cmidrule(r){1-4}
    Individual & Intervention & Health Care & - Behavioral Cognitive Therapy \\
     &  & & (Woebot, Todaki\cite{Jang2021})\\
     &  & & - Promotion for Mental Wellbeing (Shim)\cite{Ly2017} \\
     &  & & - Cancer Cares for Young Survivors \\
     &  & &  (Vivibot) \cite{Greer2019} \\
\cmidrule(r){1-4}
    Individual & Intervention & Workplace & - A Chatbot for Improvement for Sedentary     \\
     &  & & Behaviour and Wellbeing (Welbot) \cite{Haile2020} \\
\cmidrule(r){1-4}
    Group & Analysis & Emotion Analysis & - Images \cite{Tan2017}\\
     &  & & - Sounds \cite{Franzoni2020}\\
     &  & & - Videos \cite{Luque2020} \\
     &  & & - Social Media \cite{gong2019} \\
\cmidrule(r){1-4}
    Group & Intervention & Mental Health & - A Chatbot for Positive Messaging \\
     &  & & (Sunny) \cite{Narain2020}  \\
\cmidrule(r){1-4}
    Group & Intervention & Group Discussion & - Group Discussion Facilitation \\
         &  & Facilitation & such as BlahBlahBot \cite{Shin2021}, \\ 
         &  & & Micbot \cite{Tennent2019}, Groupfeedbot \cite{Kim2020}, \\
         &  & & Vulnerable-Robots \cite{Traeger2020} \\
    \bottomrule
  \end{tabular}
   \caption{A list of examples for analysis and interventions for wellbeing-related factors in different categories.}
  \label{tab:table1}
\end{table}

Detection and intervention of group wellbeing with AI are not well studied. However, related studies on automated group emotion and artificial agents for social interactions have been active recently.

Automated group-level emotion recognition has been studied recently since 2012 (Table. \ref{tab:table1}) \cite{Veltmeijer2021}. Group emotion is not a simple sum of individual emotions in a group. Instead, automated emotion recognition needs to track unique group emotion dynamics. A user survey has been developed as a proxy of such group emotion \cite{Dhall2015}. Multiple studies predict group emotional labels based on various datasets from images \cite{Tan2017} videos \cite{Luque2020}, and social media \cite{gong2019}. Such studies target different sizes and states of seated, standing, and dynamic groups. Veltmeijer et al. pointed out three technical challenges \cite{Veltmeijer2021}. First, group size changes. Second, subgroup emotions in a larger group can be different. Third, group emotion can also change. Although methods are under development, automated emotion detections for groups are perhaps applicable to group wellbeing detection. Several types of researches, applications, and commercial products for interactions with AI have been introduced in various situations such as education, hospitals, games, workplaces, social media, banks, online dating, sports, tourism, etc. These agents are expected to increase learning speed, team performance, successful dating matches, or satisfaction during traveling.

Not only AI without agents but also AI-augmented agents are widely used in our societies. We define such artificial agents as three types: robots, social bots, and avatars. AI-augmented agents are commonly used for cooperative purposes for interactions between human and artificial agents. Human-robot studies are commonly done to understand the capability of robots \cite{Sheridan2016} and how humans recognize robots \cite{Chae2016,Lucas2014}. Social bots have also been studied for communications through apps and social media \cite{Assenmacher2020}. Avatar-human interactions are further studied in the context of remote learning although humans control such avatars in most studies \cite{Chae2016} These studies aim for interactions between human and artificial agents.

An emergent application of artificial agents as social mediators is expected to promote social interactions between humans and prevent problematic behaviors within a group \cite{Chita-Tegmark2020,Dafoe2021,Rahwan2020}. However, much fewer studies target social groups for wellbeing. Some recent studies worked on discussion facilitation with artificial agents \cite{Kim2020,Traeger2020,Tennent2019} and wellbeing promotion \cite{Narain2020}. The group intervened in a social group to induce conversations and engagement on problem-solving. One study with a messenger chatbot, Sunny, worked on group wellbeing by sending positive messages to 4 member groups and had positive effects on wellbeing \cite{Narain2020}. These studies are mostly limited in discussion facilitation, but potential applications of social groups can be more extensive in different fields such as houses, schools, hospitals, caregivers, sports, workplaces, social media, tourism, where social groups are formed. For example, artificial agents could work for team engagements in sports by giving analysis or feedback based on their performance. The AI-augmented agents could also work on the mediation of conflicts between members in workplaces as well as enhancement of discussion facilitation. In doing so, we may expect artificial agents to promote wellbeing.

It is also critical to ask whether we explicitly need such artificial agents. It may be sufficient to have AIs without agents such as recommendation systems for e-commerce. One benefit of artificial agents could be related to attentional engagements by agents \cite{Chae2016,Lucas2014,Mollahosseini2018,Spicer2021}. Multiple studies showed artificial agents enhance engagements \cite{Oertel2020}. This attentional engagement can be augmented by the appearance of artificial agents \cite{Li2010,Bente2008}. Several pieces of evidence also revealed that the appearance of artificial agents influences human trust of the agents and induces similar human behaviors to humans by the agents \cite{Caruana2017,Lucas2014}. It is uncertain whether artificial agents work best in all situations, but they may exert stronger influences than just non-agentive AIs via emulating human-like interactions.

In sum, previous studies on the analysis of emotional detection for social groups and intervention of social interactions are active. However, these analyses and interventions on social interactions have not yet merged, and few studies focus on group wellbeing.

\subsection{Communicating with Social Human Groups via Artificial Agents}

The mediative role for human-human interactions with artificial agents has not been well studied. Potential opportunities of such artificial agents are more extensive than current opportunities for individual wellbeing. However, the mediative role of AI in group wellbeing, the so-called relational approach, has not been explicitly explored. Here, we clarify two types of a relational approach to social groups. By doing so, we prompt the development of relational approaches for group wellbeing.

\begin{figure}
  \centering
  \includegraphics[scale=0.6]{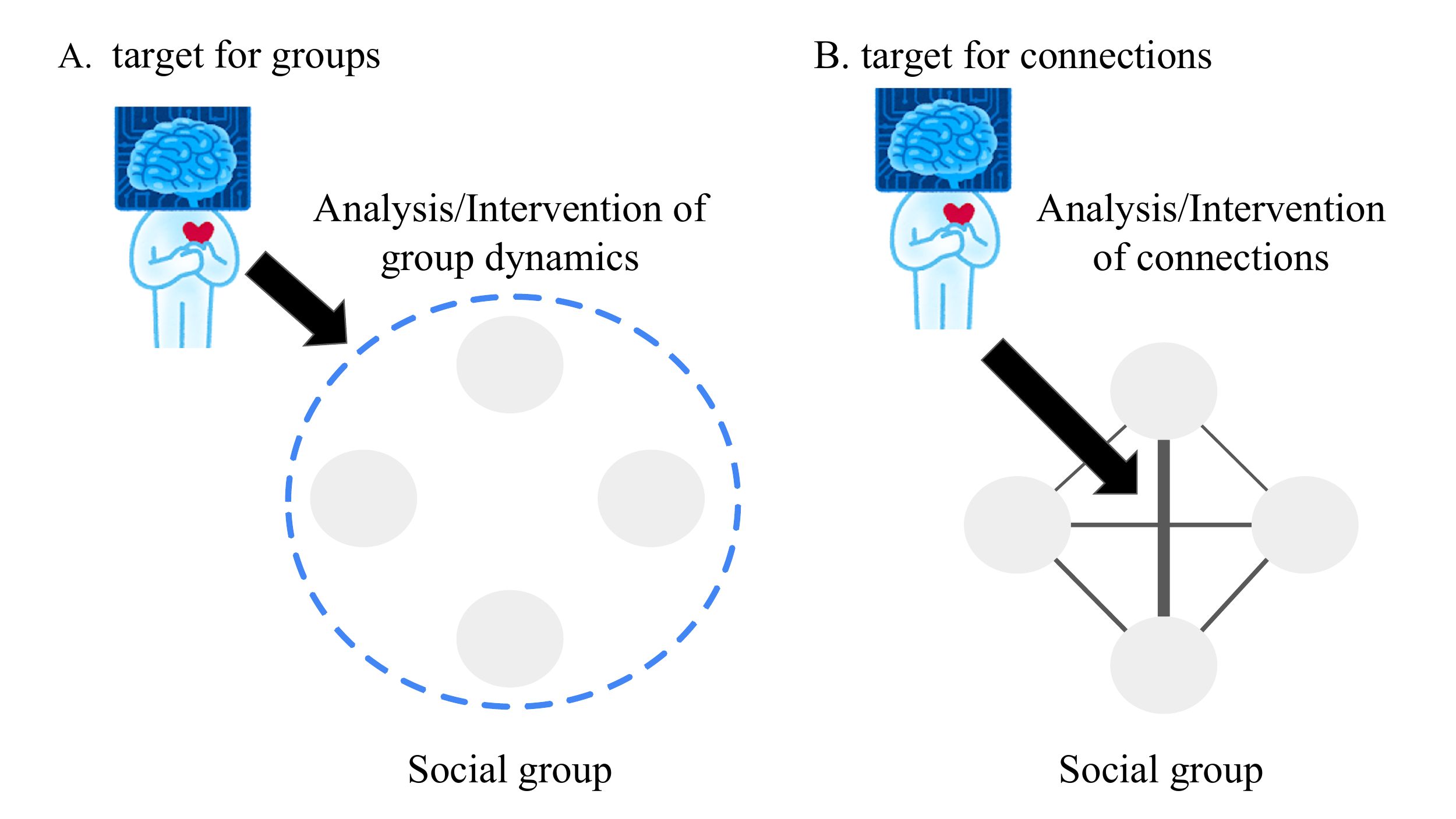}
  \caption{Intervention on group dynamics. A. AI agents analyze group dynamics, and intervene a social group by communicating to all members. B. AI agents analyze one-to-one interactions within a social group and intervene social connections or specific members based on member’s connections.}
  \label{fig:fig1}
\end{figure}

One type of the relational approach is to analyze group dynamics itself from conversations or their behaviors (Figure.\ref{fig:fig1}.A). The previous studies on automated emotion detection target analyzing such group dynamics by facial expressions and conversations \cite{Kim2020,Narain2020,Tennent2019,Traeger2020}. A robot agent study also targets group performance such as total conversation time (vulnerable-robots)\cite{Traeger2020}. This approach focuses on average or wholistic group dynamics not considering the relationship among members in a group.

Another type of the relational approach is to analyze one-to-one member interactions in detail (Figure.\ref{fig:fig1}.B). We also directly contact a person in management not by groups. For example, when certain group members engage in group discussion, another member familiar with one of the members may have insight. In this case, you as a mediator want to ask the member to promote his or her engagement in the discussion. Such a role can be served by an artificial agent. This type should be computationally intensive since N-to-N human interactions are analyzed based on methods such as computer vision and natural language processing \cite{Poria2019}. Along with such development, computation power increases these days, so current computation power could be sufficient for human-human interactions within a few members.

It is interesting to ask whether these two types of the relational approaches can be integrated as computational methods like social network analysis \cite{Gesell2013} and network controllability \cite{Liu2011}. One potential key field is related to network analysis considering both each social connection and network organization. In neuroscience, analysis and intervention of whole-brain state with region-region interactions are actively studied \cite{Tang2018}. Such network analysis further would bring an integrative perspective of social interactions and group dynamics. These approaches should be enriched by the development of the two types of the relational approach for group wellbeing. The distinction could be tentative but should be useful to work on wellbeing AI from a view of social interactions.

\section{Challenges on Relational Approach and Potential Ethical Issues}
We discuss three types of challenges on designing and managing wellbeing AI for the human-human social interactions based on the relational approach. 1) Changes related to fairness issues of computation and authority from the viewpoint of different cultural contexts, conflict of interest, and structure of benefits. 2) Challenges related to the privacy of human-human interactions from the viewpoint of ownership and autonomy of communications. 3) Challenges related to usefulness from the viewpoint of users of accessibility and safety. These challenges must be overcome for the effective and successful introduction and management of wellbeing AI.

Fairness of computation is raised as an important issue in AI research. Fairness in AI research is composed of three perspectives: fairness, conflict of interest, and respect of different communities. First, each social connection within a group should be fairly considered. Asymmetrical social connections may cause issues within a group. Next, the introduction of wellbeing AI by administrators should be fairly considered for users. Wellbeing AI can be expected by administrators to enhance engagements of users in workplaces, social media and games. Such increased engagements have the potential to deteriorate life satisfaction causing burnout symptoms in the long term. Long-term wellbeing for users, then, should be considered. Third, different cultures of communities should be respected. Perception of wellbeing is known to differ in different communities and populations \cite{Lai2013}. This example may reflect individual traits based on experience, personality, and genetics. Computation of social connections should not only consider a specific type of individual traits but also multiple perceptions.

Privacy of human-human interactions is another crucial issue. Multiple issues have been raised by previous actions by companies on controlling and using the private data of users. In this regard, privacy, autonomy, and ownership of social interactions should be considered. First, excessive access and storage of private data should not be permissible. Communications are performed with image, auditory, and text information but storing, analyzing, and providing to a third party should depend on the permission of users. Such actions potentially causing disadvantages to users should not further be taken. Second, whether actions by wellbeing AI are excessive should be considered. Such actions may cause behavioral constraints for users. Such interventions to corruption of autonomy in groups should not be permissible. Third, appropriate interventions on social connections should be considered. Related to autonomy preservation, some interventions may be permissible depending on group characteristics, but others not. This is perhaps related to a discussion of moral agency in AI where what AI agents are allowed to perform. Members within groups should determine which type of analysis and interventions is permissible.

Whether interventions to users are appropriate or not is critical. One reason is designing wellbeing environments can be more important than introducing wellbeing AI. For example, the relationship between employees and wellbeing is dependent on working environments. In other words, it might not be important to have such wellbeing AI if the working environment is not designed to promote wellbeing. This idea might be aligned with wellbeing in the importance of game design rather than the importance of introducing wellbeing AI. Appropriateness of wellbeing should be considered from usability, understanding, and the public interest of users. The stability of wellbeing AI is the priority to be considered. Attempts to have wellbeing AI is still under exploration. Real applications might not often be welcomed in social contexts. What factors determine such usability for users should be investigated. Second, understanding users is important. Related to usability, the mismatch between users and applications might be associated with misunderstanding of users by administrators. Third, a perspective of public interest is needed. This is a third-party view of wellbeing AI. Even though users and administrators gain benefits from applications of wellbeing AI, the relational approach may have a huge harmful impact on the public interest. Such appropriateness should be considered too.

Multiple challenges including three types of perspectives exist for designing and managing the relational approach of wellbeing AI since such approach is implicitly under development. Nonetheless, the relational approach of wellbeing AI has huge room to benefit our societies.

\section{Conclusion}
In this paper, we introduced the notion of AI-supported wellbeing in the era of digital worlds and presented an overview of the relational approach to promoting positive social interactions by analyzing and managing human-human interactions for the introduction of AI-supported wellbeing in the era of digital worlds. First, we described psychological research on wellbeing based on personality, genetics, and behavioral and cognitive interventions, and concluded that social relationships are crucial for wellbeing. We, then, identified an unexplored category of wellbeing AI and group wellbeing. Group wellbeing through telecommunications is especially critical since the expansion of telecommunications may cause psychological issues such as distress and loneliness which are reported during COVID-19.

By reviewing previous literature of interventions on social networks with a robot and virtual agents, we further introduced the relational approach, which analyzes and mediates human-human interactions with artificial agents such as chatbots and robot agents. The relational approach mediates human-human social interactions in the real or digital world to promote wellbeing and other factors such as team performance. Finally, we discussed potential challenges of design and usage of the relational approach in wellbeing AI to establish its successful support of human social networks. We shed light on the mediative roles of AI-augmented agents to benefit human mental health and wellbeing in real and digital environments. By doing so, we expect a broader understanding and further development of group wellbeing.

\section{Acknowledgments}
We acknowledge Toshifumi Sasaki in Osaka University for helpful discussion.

\bibliographystyle{unsrt}  
\bibliography{references}

\end{document}